\documentclass[twocolumn,showpacs,preprintnumbers,amsmath,amssymb]{revtex4}
\usepackage{subfigure}
\usepackage{graphicx}

\begin{document}

\newcommand{\bc}{\begin{center}} \newcommand{\ec}{\end{center}}
\newcommand{\be}{\begin{equation}} \newcommand{\ee}{\end{equation}}
\newcommand{\beqn}{\begin{eqnarray}} \newcommand{\eeqn}{\end{eqnarray}}

\title{Finite-size scaling of pseudo-critical point distributions\\ 
in the random transverse-field Ising chain}

\author{Ferenc Igl\'oi}  
\affiliation{ Research Institute for Solid
State Physics and Optics, H-1525 Budapest, P.O.Box 49, Hungary}
\affiliation{ Institute of Theoretical Physics, Szeged University,
H-6720 Szeged, Hungary}
\author{Yu-Cheng Lin and Heiko Rieger}
\affiliation{
 Theoretische Physik, Universit\"at des Saarlandes, D-66041
 Saarbr\"ucken, Germany }
\author{C\'ecile Monthus}
\affiliation{Service de Physique Th\'eorique,
  Unit\'e de recherche associ\'ee au CNRS, \\
  DSM/CEA Saclay, 91191 Gif-sur-Yvette, France}
\date{\today}

\begin{abstract}
We study the distribution of finite size pseudo-critical points in a
one-dimensional random quantum magnet with a quantum phase transition
described by an infinite randomness fixed point. Pseudo-critical
points are defined in three different ways: the position of the
maximum of the average entanglement entropy, the scaling behavior of
the surface magnetization, and the energy of a soft
mode.  All three lead to a log-normal distribution of the
pseudo-critical transverse fields, where the width scales as
$L^{-1/\nu}$ with $\nu=2$ and the shift of the average value scales as
$L^{-1/\nu_{typ}}$ with $\nu_{typ}=1$, which we related to the scaling
of average and typical quantities in the critical region.
\end{abstract}


\maketitle

Quenched disorder has a profound effect on the physical
characteristics of phase transitions in classical and quantum mechanical
systems. A theoretically and experimentally important issue is the
measurement of physical observables in disordered systems at or near
critical points. These measurements are always performed on finite
samples and on one particular realization of disorder. Finite size
scaling (FSS) \cite{fss} is the systematic way to extract informations
on the thermodynamic limit by studying finite systems, and the objects
to be analyzed by FSS of disordered systems are the distributions of
physical properties in the ensemble of the disorder realizations. This
also sheds light on the question about whether a single experimental
measurement on a rather large system is representative for the whole
ensemble of random systems to which it belongs. This is very much
connected to the important issue of self-averaging
\cite{domany,aharony} of thermodynamic quantities like
the expectation values for order parameter, specific heat or
susceptibilities. 

In an infinite system these observables display a characteristic
singularity at a critical point, where for instance the susceptibility
diverges. This divergence is suppressed in a finite system and
replaced by a finite maximum, the location of which is called
pseudo-critical point and is slightly shifted against the infinite
system's critical point. In pure systems this shift depends in a
systematic way on the lateral systems size $L$, usually proportional
to $L^{-1/\nu_P}$, where $\nu_P$ is the correlation length exponent of
the pure system. In finite disordered systems the susceptibility
usually has several maxima in the critical region and each one is
slightly shifted against the critical point of an infinite system. One
identifies the location of the largest maximum with the pseudo-critical
point of the corresponding sample and an intriguing question therefore
concerns the distributions of these pseudo-critical points.

If the disorder is irrelevant according to the Harris criterion
$\nu_P>d/2$ \cite{harris}, $d$ being the dimension of the variation of
the disorder (usually identical with the system's spatial dimension),
the width of the distribution scales as $L^{-d/2}$. In this case the
shift of the average finite-size transition point is proportional to
$L^{-1/\nu_P}$ as the shift of the finite size pseudo-critical point in
the pure case.

For relevant disorder $\nu_P<d/2$, there is a new random (R) fixed
point at which the exponent, $\nu_R \ne \nu_P$, satisfies the
relation\cite{ccfs} $\nu_R \ge d/2$. In many random systems (diluted
magnets, random field problems, spin glasses, etc.) the random fixed
point is of conventional form, which means that thermal and disorder
fluctuations remain of the same order at large scales, i.e. during
renormalization.  According to the finite-size scaling theory of
conventional random critical points \cite{aharony} both the shift and
the width of the distribution of pseudo-critical points are characterized by
the same random exponent and there is a lack of self-averaging
\cite{domany}. These predictions were checked for 
various models \cite{domany,aharony,paz2,mai04,PS2005}.

In this paper we intend to examine the distribution of pseudo-critical
points and its finite size scaling for a quantum phase transition in a
paradigmatic model for a random magnet with an infinite randomness
fixed point. This new type of random fixed points has been observed in
various systems (disordered quantum magnets at $T=0$, random
stochastic models, etc.) in which the disorder plays a dominant r\^ole
over quantum, thermal, or stochastic fluctuations : during
renormalization the strength of disorder grows without
limits\cite{fisher}. Many asymptotically exact results have been
obtained for one dimensional systems, partially by the use of a strong
disorder renormalization group (SDRG) method \cite{im}.

We study in detail the random transverse-field Ising spin
chain\cite{mccoy} (RTFIC) defined by the Hamiltonian:
\be
H=-\sum_l J_l \sigma_l^x \sigma_{l+1}^x-\sum_l h_l \sigma_l^z\;,
\label{hamilton}
\ee
in terms of the $\sigma_l^{x,z}$ Pauli matrices at site $l$. Here the
$J_l$ exchange couplings and the $h_l$ transverse-fields are
independent random variables. We are interested in the properties
of the system in its ground state, i.e.  at $T=0$. In the
thermodynamic limit the control-parameter is given by\cite{pfeuty}
\be
\delta=[\ln h]_{\rm av}-[\ln J]_{\rm av}\;,
\label{h_infinity}
\ee
where $[\dots]_{\rm av}$ denotes averaging
over quenched disorder, so that $\delta=0$ at the critical point\cite{pfeuty}. 
In the following, we consider the case of random couplings $J_l$
and homogeneous transverse fields $h_l=h$ \cite{note1}, so
$h$ is the analog of the temperature in thermal transitions. 
Its critical value in the thermodynamic limit is given by $\ln h_c(\infty)=[\ln J]_{\rm av}$.
In the vicinity of the critical point the average correlation length
involves the exponent $\nu=2$, whereas the typical correlation length
diverges with a different exponent $\nu_{typ}=1$
\cite{fisher}. The characteristic time-scale, $\tau$, which is related
to the smallest gap as $\tau \sim \epsilon^{-1}$, scales
logarithmically at the critical point, $\ln \tau \sim
\sqrt{L}$. It remains divergent also in an extended
region of the off-critical region, in the so called Griffiths phase,
where $\tau \sim L^{z}$ with a dynamical exponent, $z$, which depends
on the distance from the critical point.

Finite-size scaling of the RTFIC has been studied in
\cite{bigpaper,dharyoung,microcano}. The
distribution of the surface magnetization could be computed analytically 
\cite{dharyoung,microcano}, and the distribution of the 
the gap and the end-to-end correlation function by the SDRG method
\cite{microcano}. They turned out to be different in the
micro-canonical and the canonical ensembles. Here we define sample
dependent critical parameters $h_c(\alpha,L)$ (where $\alpha$
indicates a particular disorder realization) and study their
distribution. The standard approach, defining $h_c(\alpha,L)$ through
the rounding of the singularity of the susceptibility, does not work
here, since the susceptibility is divergent also in the Griffiths
phase. A similar problem arises for the rounding of the specific heat,
since it has only a very weak essential singularity.

\paragraph{Pseudo-critical points through the maximum of the average entropy}
Here we suggest that for random quantum systems the pseudo-critical
transition points can be conveniently defined through the rounding of
the average entanglement entropy. For the RTFIC we consider a periodic
sample $(\alpha)$ of length $L$ and calculate the entanglement entropy
between the two halves of the chain, which is then averaged over all
possible starting points of the block. In the limit $L\to\infty$ the
average entropy is divergent at the critical point\cite{refael},
whereas in a finite sample we use the position of its maximum to
define $h_c(\alpha,L)$.

\begin{figure}[t]
\begin{center}
\includegraphics[width=3.2in,angle=0]{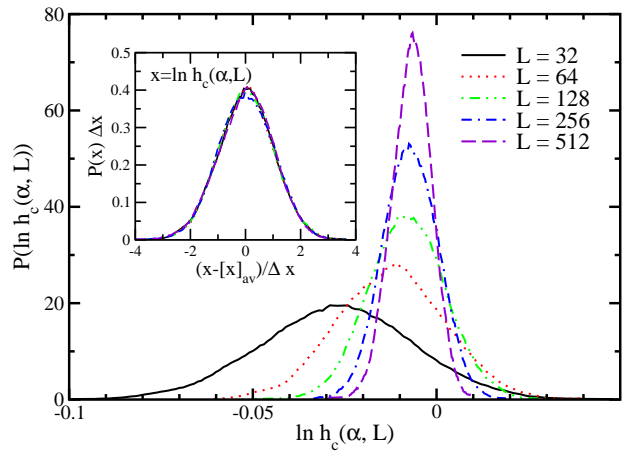}
\end{center}
\caption{
\label{fig:distr} (Color online)
Distribution of $\ln h_c(\alpha,L)$ for a disorder with $\Delta J=0.4$ for
different sizes. In the inset the distributions of scaled variables
is well described by a Gaussian.}
\end{figure}

\begin{figure}[h]
\begin{center}
\includegraphics[width=3.2in,angle=0]{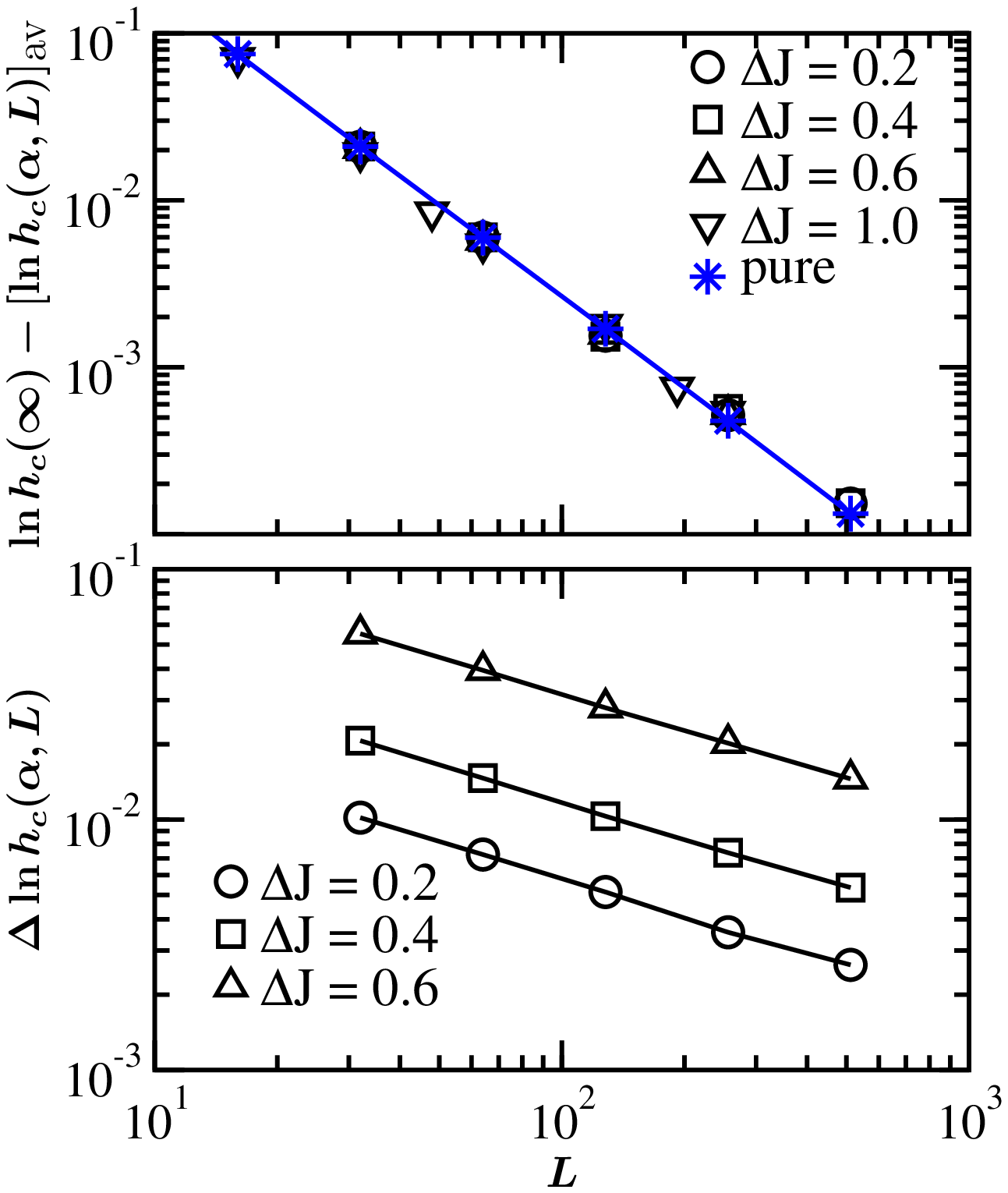}
\end{center}
\caption{
\label{fig:shift} (Color online)
Upper panel: Shift of the average of $\ln h_c(\alpha,L)$ as a function
of $L$ in a log-log scale. The data points have practically no
disorder dependence. The straight line has a slope of $1.8$, which
represents an effective exponent close to 2. Lower panel: Scaling of
the width of the distribution of $\ln h_c(\alpha,L)$ as a function of
$L$ in a log-log scale. The slope of the straight lines is practically
disorder independent: $1/\nu=0.50(1)$ }
\end{figure}

In the numerical calculations we have used efficient free-fermion
techniques \cite{lsm,peschel} with which we could calculated the 
entanglement entropy up to sizes $L=512$. The couplings in
Eq.(\ref{hamilton}) are taken from a uniform box-like distribution,
which is centered at $J=1$ and has a width $\Delta J$. For different
strengths of disorder ($\Delta J=0.2,~0.4,~0.6$ and $1.0$) and for
each system size $L$ $10000$ disorder realizations were generated.
Additionally the entropy of each sample is
averaged over $L/2$ starting position of the block. As an
illustration we show in Fig.\ref{fig:distr} the probability
distributions of $\ln h_c(\alpha,L)$ at a disorder $\Delta J=0.4$ for
different sizes.  The distribution functions for different $L$ are
symmetric and in terms of rescaled variables: $(\ln h_c(\alpha,L)-[\ln
h_c(\alpha,L)]_{\rm av})/\Delta \ln h_c(\alpha,L)$ the transformed
distributions are well fitted by the same Gaussian form, as shown in
the inset of Fig.\ref{fig:distr}. For different strength of disorder
we have analyzed the shift of the average value, $[\ln
h_c(\alpha,L)]_{\rm av}$, as well as the standard deviation, $\Delta
\ln h_c(\alpha,L)$, which are shown in
Fig.\ref{fig:shift}. Interestingly, the average transition point for a
given $L$ is practically independent of the strength of disorder and
corresponds to the value in the pure system. 

Our numerical data are compatible with a FSS form for the shift that is
given by $h_c(\infty)-h_c(L) \sim L^{-2}ln L$. The scaling of the
width of the distributions is found to follow $\Delta h_c(L) \sim
L^{-1/\nu}$, where the exponent is given by $1/\nu=0.50(1)$
independently of the strength of disorder (Fig.\ref{fig:distr}). Thus,
our numerical estimate for $\nu$ agrees well with the exponent of the
average correlation length of the RTFIC. We can thus conclude that the
distribution of $\ln h_c(\alpha,L)$ in Fig.\ref{fig:distr} is well
described by a Gaussian with a variance of $O(1/L)$ and by a shift of
$O(L^{-2}ln L)$.


\paragraph{Pseudo-critical points through the surface magnetization}
The surface magnetization is perhaps the simplest physical quantity of the RTFIC. Fixing the spin at one
end of the chain say at $l=L+1$, which amounts to have $h_{L+1}=0$, the magnetization at the other
end of the chain at $l=1$ is given by the exact formula\cite{mspeschel}:
\be
m_s=[1+\textstyle\sum_{l=1}^{L} \textstyle\prod_{j=1}^l
(h_j/J_j)^2]^{-1/2}\;.
\label{peschel}
\ee
In the following we use a doubling procedure\cite{PS2005}: For a given random sample 
($\alpha$) of length $L$, we construct a replicated sample ($2\alpha$) of length
$2L$ by gluing two copies of ($\alpha$) together, and study the ratio of the 
surface magnetizations: $r(\alpha,L)=m_s(2\alpha,2L)/m_s(\alpha,L)$.
We rewrite the exact expression in Eq.(\ref{peschel}) as
\begin{equation}
m_s(\alpha,L) = \left[ Z_1(L)+1 \right]^{-1/2},\quad
Z_1(L)=  \sum_{l=1}^{L} e^{ - U(l) }\;,
\end{equation}
in terms of a random walk variable: $U(i)= 2 \sum_{j=1}^i \ln \frac{ J_j}{h_j}$.
Similarly, we obtain for the replicated sample
\begin{equation}
m_s(2\alpha,2L) = \left[ Z_2(L)+1 \right]^{-1/2},\quad  Z_2(L) =  \sum_{l=1}^{2 L} e^{ - U_{2}(l) }
\label{Z_2}
\end{equation}
where $U_2(i)= U(i)$ for $1 \leq i \leq L$ and it is 
$U_2(i)= U(L)+U(i-L)$ for $L+1 \leq i \leq 2L$.
The expression in Eq.(\ref{Z_2}) simplifies into
$Z_2(L)= \left(1+e^{-U(L)} \right) Z_1(L)$. Since in the critical
region $\ln Z_1(L) \sim L^{1/2}$ for large $L$ the ratio
of the two surface magnetizations is given by:
\begin{equation}
 r(\alpha,L) \simeq \left(1+e^{-U(L)} \right)^{-1/2}
\label{m2m_m}
\end{equation}
For the pure chain in the ordered phase with
$\delta_P=h-h_c(\infty)=h-1<0$ we have $U(L) \to \infty$, and in a
large finite system the ratio behaves as $r(L)=1-e^{-2|\delta_P|L}/4$,
whereas in the disordered phase with $\delta_P>0$ we have $U(L) \to
-\infty$ and the finite-size correction reads as $r(L)=e^{-\delta_P
L}$. Then at the critical point $U(L)=0$ and the ratio has a
non-trivial value, $r(L)=1/\sqrt{2}$. For the random chain we have the
same type of trivial fixed points corresponding to the disordered
($U(L) \to \infty$) and the ordered ($U(L) \to -\infty$) phases,
respectively, and it is natural to use the condition, $U(L)=0$, to
define finite-size critical transverse fields. This leads to the
micro-canonical condition:
\be
\frac{1}{L}\sum_{l=1}^L \ln J_l= \frac{1}{L}\sum_{l=1}^L \ln h_l = \ln h_c(\alpha,L)\;.
\label{microcan1}
\ee
Using this definition we obtain for the finite-size behavior in the
disordered phase for a given sample $\ln m_s(\alpha,L) \sim - (\ln
h_c(\alpha,L)-\ln h)L$ and for its average: $[\ln m_s(\alpha,L)]_{\rm
av} \sim - (\ln h_c(\infty)-\ln h)L\sim - |\delta|L$, which involves
the typical exponent, $\nu_{typ}=1$.

\paragraph{Pseudo-critical points through a soft mode}
Criticality of the RTFIC in the free fermion representation is related to
the vanishing of the excitation energy of a special fermionic mode. We
recall that for closed chains, i.e. with $J_L \ne 0$ the even and odd
number of excitations are taken from two different sectors and the
ground state is the vacuum of the even sector, whereas the first
excited state is in the odd sector and contains one fermion with
energy, $\Lambda_{1}$.  For the pure system its energy is given by
$\Lambda^P_{1}=h-J=\delta_P$, thus it changes sign at the critical
point.  Here we use the same condition, $\Lambda_{1}=0$, to define
criticality a in finite random system, too. This leads to the matrix
equation\cite{lsm}:
$(\mathbf{A}+\mathbf{B})\mathbf{\Phi}_{1}=\mathbf{0}$ with
\be
(\mathbf{A}+\mathbf{B})=
\begin{pmatrix}
h_1  & J_1 &      &       &       &       &       \cr
     & h_2 & J_2  &       &       &       &       \cr
     &     & h_3  & J_3   &       &       &       \cr
     &     &      &       &\ddots &\ddots &       \cr
     &     &      &       &       &h_{L-1}&J_{L-1}\cr
J_{L}&     &      &       &       &       &h_{L}  \cr
\end{pmatrix}
\ee
the solution of which is just the micro-canonical condition in
Eq.(\ref{microcan1}).

In the following we study the scaling behavior of the smallest gap
$\epsilon(\alpha,L)$ in the ordered phase. For this it
is more convenient to use open chains with $L$ bonds where 
we have the asymptotic expression\cite{bigpaper}:
\be
\epsilon(\alpha,L) \sim m_s(\alpha,L) \overline{m_s}(\alpha,L) \prod_{i=1}^{L} \frac{h_i}{J_i} h_1\;.
\label{epsilon}
\ee
provided the scaled gap, $\epsilon(\alpha,L)L$, goes to zero.  Here we
take $h_{L+1}=h_1$. $m_s(\alpha,L)$ and $\overline{m_s}(\alpha,L)$
denote the surface magnetization at the two ends of the chain, which
are both $O(\delta^{1/2})$ in the ordered phase. Consequently for a
given sample, $\ln \epsilon(\alpha,L) \sim - (\ln h-\ln
h_c(\alpha,L))L$ and for its average: $[\ln \epsilon(\alpha,L)]_{\rm
av}
\sim - (\ln h-\ln h_c(\infty))L \sim - \delta L$. Thus the finite-size correction involves the
typical exponent, $\nu_{typ}=1$.

\paragraph{Log-normal distribution}
We have used three different methods to determine pseudo-critical
points, all of which gave coherent results for the pseudo-critical
point distributions. We found that the distribution of $x= \ln
h_c(\alpha,L)$ is Gaussian around 
$\overline{x}=[\ln h_c(\alpha,L)]_{\rm av}$
\begin{equation}
 P_L( x= \ln h_c(\alpha,L) ) = \sqrt{
\frac{L}{2 \pi \sigma^2 }} e^{ - \frac{L}{2\sigma^2} (x- \overline{x})^2}
\label{plogh}
\end{equation}
Thus $\Delta \ln h_c(\alpha,L) = \sigma/{\sqrt L}$ and the
fluctuations are governed by the exponent $\nu=2$. The shift of the
average, $\ln h_c(\infty)-\overline{x}$, is of $O(L^{-2}\ln L)$ from
the average entropy and zero by the other two methods. The average of
the critical transverse fields is given by: $[h_c(\alpha,L)]_{\rm
av}=e^{\overline{x}}e^{\sigma^2/2L}=h_c(\infty)(1+2\sigma^2/L+\dots)$,
thus $[h_c(\alpha,L)]_{\rm av}-h_c(\infty)=\lambda(L) \sim 1/L$, which
is consistent with $\nu_{typ}=1$. We can thus conclude that $\Delta
h_c(L)/\lambda(L)$ tends to infinity, which can be taken as a
definition of an infinite randomness fixed point.

We now use these results to explain the role of the averaged
correlation length $\delta^{-2}$.  In the disordered phase
($h>h_c(\infty)$), the average surface magnetization $[m_s(h,L)]_{\rm
av}$ is dominated by the rare ordered samples having
$m_s(\alpha,L)=O(1)$, for which $h_c(\alpha,L)>h>h_c(\infty)$.  From
their probability we obtain:
\begin{eqnarray}
[m_s (h,L)]_{\rm av}  \sim Prob( \ln h_c(\alpha,L) > \ln h)
\sim e^{- \frac{L}{2\sigma^2} \delta^2} 
\label{msav}
\end{eqnarray}
Its exponential decay thus involves the critical exponent $\nu=2$.  In
the ordered phase ($h<h_c(\infty)$) we consider the average gap,
$[\epsilon(h,L)]_{\rm av}$, which is dominated by those rare
realizations, for which $h_c(\alpha,L)<h<h_c(\infty)$ and
$\epsilon(\alpha,L)=O(1)$.  From their probability we obtain:
\begin{eqnarray}
[\epsilon (h,L)]_{\rm av}  \sim Prob( \ln h_c(\alpha,L) < \ln h)
\sim e^{- \frac{L}{2\sigma^2} \delta^2} 
\label{gapav}
\end{eqnarray}
which also involves the critical exponent $\nu=2$. We note that
previously we have shown that the typical quantities, both the surface
magnetization and the gap, have an exponential decay:
$e^{-L|\delta|}$, which involve the typical exponent, $\nu_{typ}=1$.

Our results for the RTFIC are also relevant for other systems
displaying an infinite randomness fixed point, like the random
antiferromagnetic $XX$ chain and various stochastic models with
quenched disorder, such as the Sinai-walk and the partially asymmetric
exclusion process. Also in higher dimensional realizations of infinite
randomness fixed points, like the two-dimensional random transverse
Ising model \cite{motrunich,pich}, we expect a similar scenario as we
described here, which can be checked for instance by calculating the
entropy numerically by the SDRG method \cite{lin-new}.

This work has been supported by the National Office of Research and
Technology under Grant No. ASEP1111, by a German-Hungarian exchange
program (DAAD-M\"OB), by the Hungarian National Research Fund under
grant No OTKA TO48721, K62588, MO45596. F.I. thanks SPhT Saclay for
hospitality.


\begin{thebibliography}{}
\bibitem{note1} 
  Generally we can use the parametrization, $h_l=h \kappa_l$, and in this
  case $\ln J_l$ is replaced by $\ln J_l-\ln \kappa_l$.
\end{thebibliography}
\end{document}